\documentclass[twocolumn,showpacs,preprintnumbers,amsmath,amssymb]{revtex4}

\usepackage{bm}
\usepackage{latexsym}

\begin{document}
~~~~~~~~~~~~~~~~~~~~~~~~~~~~~~~~~~~~~~~~~~~~~~~PHYSICAL REVIEW
\\

$\!\!\!\!\!\!\!\!\!\!\!\!\!\!$D {\bf 76} 104040 (2007)
\\

\title{Helical Symmetry in Linear Systems}

\author{Ji\v r\'{\i} Bi\v c\'ak${}^{1,2}$ and Bernd G. Schmidt${}^2$\\
${}^1$Institute of Theoretical Physics, Faculty of Mathematics and Physics\\
Charles University, Prague, Czech Republic \\
${}^2$Max-Planck-Institut f\"ur Gravitationsphysik, Albert-Einstein-Institut\\
Am M\"uhlenberg 1, D-14476 Golm, Germany}

\begin{abstract} We investigate  properties of solutions of
the scalar wave equation and Maxwell's equations on Minkowski space
with helical symmetry. Existence of local and global solutions with
this symmetry is  demonstrated with and without sources. The
asymptotic properties of the solutions are analyzed. We show that
the Newman--Penrose retarded and advanced scalars exhibit specific
symmetries and generalized peeling properties.
%PACS numbers: 04.20.--q\ , 03.50.--z
\end{abstract}


\maketitle

\section{Introduction}
A beautiful theorem by Leon Lichtenstein  demonstrates the
existence of  solutions describing the circular motion of
two fluid bodies around their center of mass under the
Newtonian force of gravity \cite{li}. This solution is invariant
under the symmetry generated by the vector field
$\partial_t+\Omega\partial_\phi$, the helical symmetry.

Are there corresponding solutions in Einstein's theory of
gravity? For truly isolated systems such solutions
will not exist in General Relativity because the orbiting bodies
will generate gravitational radiation and spiral towards each
other. It is however conceivable  --- and this was  first
conjectured by Detweiler \cite{de}   --- that if there is the
right amount of incoming radiation which would just compensate the
outgoing radiation, then the bodies could stay on a circular orbit
and the spacetime would admit helical symmetry as an isometry.

There is an example of a solution with helical symmetry in
Maxwell's theory by Schild \cite{sch}: He shows that two opposite
electric point charges can move on a circular orbit if the force on
one particle is given by the Lorentz force of the  "one half
advanced plus retarded field" of the other particle. In this
solution the total energy of the field is infinite. (See also
\cite{bhs} where this is generalized to $N$ scalar point particles.)

One also expects that in a fully relativistic treatment of the
2--body problem with helical symmetry the ADM mass of the solution
will be infinite and that the solutions will not be asymptotically
flat in the usual sense at spatial or null infinity. Because of
this it  is not even  clear how one defines "helical symmetry" if
there is  no asymptotic symmetry group to characterize the
infinitesimal helical Killing field
$\partial_t +\Omega\partial_\phi$. 

\bigskip
 
Apart from this fascinating question whether Einstein's theory has
really solutions with helical symmetry in which incoming radiation
balances outgoing radiation, there is interest in the numerical
community to use data with approximate helical symmetry in the
numerical construction of  solutions of the 2--body
problem and the calculation of  the radiation emitted. This
question has lead to various both analytical and  numerical studies
of helical solutions of linear and nonlinear model problems by  J.
Friedman, R. Price, E. Gourgoulhon, their coworkers, and others
(see, e.g.,
\cite{fr} -- \cite{br}, and references therein). The primary goal
here is computing quasiequilibrium configurations of close binary
compact objects, or individual rotating neutron stars.

\bigskip

To address the problem of helical solutions in Einstein's  theory
it is absolutely necessary to have a really good understanding of
the  linear situation. The reason is that solutions  to  nonlinear
 problems are usually solved as  a limit of a sequence of
linear problems. No systematic study of linear helical solutions
 of the scalar wave equation, Maxwell's equations and
linearized gravity is available. This is the topic of the present
paper.

\bigskip

Section II studies solutions with helical symmetry -- i.e.
solutions invariant under the symmetry generated by $\partial_t
+\Omega\partial_\phi$ --  of the scalar wave equation on Minkowski
space. There is a very important point: because of the assumed 
symmetry one can study the wave equation on the space of orbits
of the symmetry. There one obtains an apparently simpler
3--dimensional problem. However, the symmetry reduced wave
equation is a second order PDE which changes its  character: it is
elliptic near the center and hyperbolic far out. No general theory
is available  in this situation. There is  the paper by Torre
\cite{tor} who obtains some existence results for the inhomogeneous
equations on part of Minkowski space.

We proceed completely differently. A lot  is known about the wave
equation in Minkowski space. In particular, the retarded and
advanced solutions of  a spatially compact source with helical
symmetry are also helical. So, with almost no work we obtain
solutions of the inhomogeneous wave equation on the space of
orbits.

The general homogeneous solution with helical symmetry can be
constructed via a spherical harmonics decomposition. So, we know
the general solution. 

\bigskip
There are helical solutions with explicit $Y_{lm}$ angular
behaviour. We give an integral representation of such solutions.

\bigskip
In Section III we investigate the asymptotic behaviour of these
solutions near spatial and null infinity. The retarded solution
of a spatially compact helical source has the usual Bondi--type 
expansion \cite{bo} at future null infinity but no  such expansion
exists at past null infinity. The standard radiation field of the
solution, as defined, e.g., in \cite{bo}, \cite{pen}, \cite{wa},
does not exist on past null infinity. Hence for the retarded plus
advance solution no radiation field is defined at all. Also, the
decay at spatial infinity is  rather weak, the field decays like
$\sin  r/ r$.
 
\bigskip
In Section IV  we calculate the retarded and advanced field of a
scalar source moving on a circle with constant velocity. The
rather complicated periodicity properties  of the solutions
become apparent.
 
\bigskip
The analogous solution in Maxwell's theory is presented in
Section V. We give the fields in terms of the null tetrad NP
(Newman--Penrose) projections, i.e., as three complex scalars.
Since the charged particles moving in circular obits occur in
synchrotrons (where they are guided by external magnetic fields)
our final expressions might be of some use also elswhere. In
Schild's work \cite{sch} the field is not given. The usual radiation
field again does not exist but the NP scalars exhibit generalized
peeling properties. As a consequence of the helical symmetry, the
retarded and advanced  NP scalars show specific symmetry properties.

\bigskip
From the results in Maxwell's theory one can almost guess the
formulae for Einstein's linearized gravity if needed.

\section{The wave equation on Minkowski space}

We begin by considering the wave equation on Minkowski
space and want to describe  global and local solutions
with helical symmetry.

A function $F(t,r,\theta,\phi)$ (in standard spherical
coordinates on Minkowski space) is invariant under
$\partial_t+\Omega\partial_\phi$ if and only if 
\begin{equation}
%  %\label{}
F(t,r,\theta,\phi)=G(\phi-\Omega
t,r,\theta)=G(\hat\phi,r,\theta)\ \ .
\end{equation}
The function $G$ is periodic in its first argument with period $2\pi.$ 
$F$ is a solution of the scalar wave equation if and only if
$G$ satisfies
\begin{eqnarray}
\label{rwe}
-\Omega^2G_{,\hat\phi\hat\phi}&+&G_{,rr}+{\frac{2}{r}}G_{,r}+
\\\nonumber
&+&{1\over r^2\sin\theta}(\sin\theta\  G_{,\theta})_{,\theta}+
{1\over r^ 2\sin^2\theta}{G_{,\hat\phi\hat\phi}}=0\ .
\end{eqnarray}
(In Secs. II and III we set the speed of light $c=1$.)
A remarkable property of this equation is that it changes
its character! The norm of the helical Killing vector is
$N=-1 +\Omega^2{r^2\sin^2\theta}$ and therefore the helical Killing
vector is timelike near the axis ($N<0)$, null for $N=-1
+\Omega^2{r^2\sin^2\theta}=0$,  and spacelike  for $N=-1
+\Omega^2{r^2\sin^2\theta}>0$. This implies that the reduced equation
(\ref{rwe}) is elliptic near the center and hyperbolic for $N=-1
+\Omega^2{r^2\sin^2\theta}>0$. This raises interesting questions
concerning solutions defined globally. We will denote the
three regions as:
$T=\{(\hat\phi,r,\theta)|N<0\}$,
$L=\{(\hat\phi,r,\theta)|N=0\}$,
$S=\{(\hat\phi,r,\theta)|N>0\}$. 

\subsection{Local source--free solutions}
Obviously in $T$ local solutions can be determined by
boundary value problems and in $S$ by Cauchy problems. What
happens near $L$ is not so clear.

Suppose we have a $ C^2$ solution of (\ref{rwe}) defined for
$r_1<r<r_2$ such that $N$ changes its sign in this region.
We can expand into spherical harmonics
\begin{equation}%%\label{}
G(\hat\phi,r,\theta)=G(r)\ Y_{lm}(\hat\phi,\theta)\ ,\ \
\hat\phi\in[0,2\pi)\ ,
\end{equation} 
and obtain for $G(r)$ the radial equation
\begin{equation}%\label{}
m^2\Omega^2G+{1\over r}(rG)_{,rr}-{1\over r^2}l(l+1) G=0\ .
\end{equation}
The general solution in terms of spherical Bessel functions
is
\begin{equation}\label{5}
G_{lm}(r)= {a_{lm}\over\sqrt{m\Omega
r}}J_{l+{1\over2}}(m\Omega
r)+{b_{lm}\over\sqrt{m \Omega
r}}Y_{l+{1\over2}}(m\Omega r)\ .
\end{equation} 
The set $L$, the light cylinder, is no singularity for the
radial equation. Hence, we see that there are local solutions
for which nothing particular happens at $L$.

The only solutions regular at the origin are given by
$b=0$. For $m\Omega r \gg l$ the solutions behave as  
$\sin(m\Omega r-{\pi\over 2}l)/{ m\Omega 
r}$.

Thus we know the general $C^2$ solutions in terms of
 converging superpositions of spherical harmonics and Bessel functions.
\begin{equation}%\label{}
G(\hat\phi,r,\theta)=\sum_{l,m}a_{lm}
G_{lm}(r)Y_{lm}(\hat\phi,\theta)\ .
\end{equation}
The coefficients $a_{lm},b_{lm}$ in (\ref{5})
may depend on $\Omega$. Taking (\ref{5}) in
the form
\begin{eqnarray}\label{6}
G_{lm}(r)=&& {\bar
a_{lm}\over (m\Omega)^l}{1\over\sqrt{m\Omega
r}}J_{l+{1\over2}}(m\Omega
r)+
\\\nonumber
&+&\bar b_{lm}(m\Omega)^{l+1}{1\over\sqrt{m
\Omega r}}Y_{l+{1\over2}}(m\Omega r)\ ,
\end{eqnarray} 
 where $\bar a_{lm},\bar b_{lm}$ are
independent of $\Omega$, we find, using the properties of the
Bessel functions, that in the limit $\Omega\to 0$
\begin{equation}\label{8}
G_{lm}(r)= {\bar
a_{lm}\over (2l+1)!!}r^l-\bar
b_{lm}(2l-1)!!{1\over
r^{l+1}}\ .
\end{equation} 
This is the general static 
$Y_{lm}$--solution of the Laplace equation.

\subsection{Global source--free solutions}

Now we assume that we have a $C^2$ solution defined on
Minkowski space. Again we can decompose into spherical
harmonics and because of the regularity at the origin only
$J$--Bessel functions appear:
\begin{equation}\label{gs}
G(\hat\phi,r,\theta)=
\sum_{l,m}a_{lm}{1\over\sqrt{m\Omega
r}}J_{l+{1\over2}}(m\Omega r)Y_{lm}(\hat\phi,\theta)
\ .\end{equation} 
This shows that a global $C^2$ solution is uniquely
determined by its values  in a neighborhood of the center.

What is the asymptotics of such solutions? From the $Y_{lm}$
decomposition it is not obvious what type of decay one can
produce by superpositions. Remember that a superposition of
not decaying plane wave solutions of the wave equation can
produce solutions with spatially compact support. 

There is an easy argument that we cannot obtain solutions
of finite energy: suppose we have a $C^2$ solution of  finite
energy;  then the energy in a finite box around some fixed point in space has
to decay in time, a general property of solutions of the wave equation. Such
a  decay contradicts helical symmetry.

In particular, there are no $C^2$  solutions with spatially compact support and
helical symmetry.

In the next section we shall see that there are local
solutions near null infinity which allow  a Bondi expansion,
but global source--free solutions have never a Bondi
expansion.

\subsection{Global solutions with sources}
Suppose we have a solution of the form (\ref{gs}) defined
near the center. In general, the series  will not converge
for all $r$. Examples are given by solutions of the
inhomogeneous wave equation with sources  invariant under the
helical symmetry. The simplest case is a point source moving
on an integral curve of the helical Killing field. In
spacetime the retarded and advanced fields of this source are well
defined outside the source. Near the center these solutions must
be of the form (\ref{gs}) but they become singular at the source
position. Further out we will again have a superposition of
now both Bessel function solutions.

The general  solution of the inhomogeneous wave equation
with a spatially compact source is given by  the sum of the
retarded solution and some global solution of the
homogeneous equation. 

If the spatially compact source has helical symmetry the
(unique) retarded and advanced solutions have also  helical
symmetry. 

A simple way to see this is to consider a
point charge as a source, as we do in detail
in Section IV, and change the "source
variables" $t',\phi'$ to $t'+\lambda$ and
$\phi'+\Omega\lambda$, $\lambda\in R$, which
represents the action of the helical symmetry
on the source. It is then easy to show, using
the explicit formulae for Li\'{e}nard--Wiechert potentials, that
the field remains the same when the "observation variables"
$t,\phi$ change to $t+\lambda$, $\phi+\Omega\lambda$. For an
extended source the superposition principle can be used. There is,
in fact, a more general theorem available showing for the wave
equation that  a symmetry of the source is inherited by the
retarded/advanced  fields. 

If we take these helical solutions, which we obtain from
general theorems about the wave equation in Minkowski
space, and consider them as solutions of the reduced
equation (2), we obtain immediately an existence theorem for
the reduced equation which, in contrast to the result by
Torre \cite{tor},  includes the center.

More precisely: The retarded and advanced  solution of the wave equation
with a spatially compact  helical source   defines a global
solution of the inhomogeneous reduced equation which is
analytic outside the source. The general solution is
obtained by adding a  global solution of the homogeneous
equation described in Section II B.

If we take a source with helical symmetry, then the difference of the retarded
and advanced solution defines a global, source--free solution with helical
symmetry on Minkowski space. This shows the existence of source--free
solutions defined  on all of Minkowski space.

We want to give an explicit representation of a helical solution
with $Y_{lm}$--behaviour:
Let $\rho$ be a helically symmetric   source of the scalar field $\Phi(t, {\bf r})$,
satisfying the inhomogeneous wave equation
\begin{equation}
   %\label{}
\Box \Phi=-4\pi\rho(t,{\bf r})\ ,
\end{equation}
where ${\bf r}=(x^i)=(x,y,z).$ The retarded solution reads
\begin{equation}
   \label{ret}
\Phi(t, {\bf r})=\int{{\rho(\bar t, {\bf \bar r})\over{|{\bf r-\bar
r|}}}}d\bar V\ ,
\end{equation}
with
\begin{equation}
   %\label{}
\bar t=t-|{\bf r-\bar r|}\ ,
\end{equation}
\begin{equation}
   %\label{}
|{\bf r-\bar r|}=\sqrt {r^2-2{\bf r \bar r} + \bar r^2}\ ,
\end{equation}
$r=|{\bf r}|=(x^ix^i)^{1\over 2}$, similarly for ${\bf\bar r}$.

We assume for the helical source the $Y_{lm}$ dependence:
\begin{eqnarray}%\label{}
\hat\rho(\hat\phi,r,\theta)&=&\rho(r)Y_{lm}(\hat\phi,\theta)=
\\\nonumber
&=&\rho(r)a_{lm}e^{im\hat\phi}
P_l^m(\cos\theta)\
,\
\
\hat\phi\in[0,2\pi)\,.
\end{eqnarray} 
The integrand  $I$ of the retarded integral becomes
\begin{equation}{}
 I =  {1\over |{\bf r-\bar r|}}\rho(\bar
r)a_{lm}e^{im[\bar\phi-\Omega( t-|{\bf r-\bar r|})]}
P_{l}^m(\cos\bar\theta)\ .
\end{equation}
So,
\begin{equation}{}
 I =  {\rho(\bar r)\over |{\bf r-\bar
r|}}a_{lm}e^{im\bar\phi}e^{-im\Omega
t}e^{im\Omega|{\bf  r-\bar r|}} P_l^m(\cos\bar\theta)\ ,
\end{equation}
or
\begin{equation}{}
 I =  e^{-im\Omega
t}\rho(\bar r)Y_{lm}(\bar\phi,\bar\theta){e^{im\Omega|{ \bf
r-\bar r|}} \over |{\bf r-\bar r|}}\ .
\end{equation}
Now we can use the following expansion in spherical harmonics 
(see \cite{jack}, Eqs.(16.18)--(16.22) with $m\Omega=k$):
\begin{equation}{}
 {e^{ik|{ \bf
r-\bar r|}} \over 4 \pi |{\bf r-\bar
r|}}=\sum_{l,m}g_l(r,\bar
r)Y^*_{lm}(\bar\theta,\bar\phi)Y_{lm}(\theta,\phi)\ ,
\end{equation}
\begin{equation}{}
g(r,\bar r)=ikj_l(kr_<)h_l{}^{(1)}(kr_>)\ ,
\end{equation}
with $r_<=min(r,\bar r), r_>=max(r,\bar r)$.
The Bessel and Hankel  functions $j_l, h_l{}^{(1)}$ can be found in
\cite{jack} [Eqs.(16.9), (16.10)] or \cite{ab}.

If we insert this expansion into the retarded integral, use the
orthogonality properties of the spherical harmonics and remember
the volume element we obtain
\begin{eqnarray}{}
\Phi &=&  e^{-im\Omega
t}Y_{lm}(\phi,\theta)
\\\nonumber
&&\times\int_0^\infty d\bar r
\bar r^2\rho(\bar r)im\Omega     j_l(m\Omega
r_<)h_l{}^{(1)}(m\Omega r_>)\ ,
\end{eqnarray}
or
$$
\Phi =  e^{-im\Omega
t}Y_{lm}(\phi,\theta)im\Omega h_l{}^{(1)}(m\Omega r)\int_0^r
\!\!\!\!\! d\bar r \bar r^2\rho(\bar r)     j_l(m\Omega \bar r)
$$
\vskip -6mm
\begin{equation}
+e^{-im\Omega
t}Y_{lm}(\phi,\theta)im\Omega j_l(m\Omega r)\!\!\int_r^\infty\!\!\!\!\!\! d\bar r
\bar r^2\rho(\bar r)     h_l{}^{(1)}(m\Omega \bar r )~.
\end{equation}
For compactly supported $\rho(r)$ the behaviour near infinity is
given  by $h_l{}^{(1)}(m\Omega r)\sim e^{im\Omega r}/ m\Omega r$.
In this representation of the solution the "retardation in
space time" is completely hidden. 

\bigskip
Thus, we have complete control of all solutions of the inhomogeneous
reduced equation with signature change  on the quotient. The
lesson is not to forget Minkowski space and just work with the
reduced equation on the quotient of the symmetry, but use all the
results one has on the wave equation in Minkowski space!

\section{Asymptotics of solutions with sources}
In this section we discuss the asymptotics of helical solutions
on Minkowski space for scalar fields, and we also mention the
analogous cases for Maxwell fields and linearized gravity. Here
we deal with general (unspecified) sources of spatially compact
support. In the next sections we investigate the fields of point
sources on circular orbits in detail.
\subsection{The scalar wave equation}
Let $\rho$ be a helically symmetric   source of the scalar field $\Phi(t, {\bf r})$,
satisfying the inhomogeneous wave equation
\begin{equation}
   %\label{}
\Box \Phi=-4\pi\rho(t,{\bf r})\ ,
\end{equation}
where ${\bf r}=(x^i)=(x,y,z).$ Again, the retarded solution reads
(\ref{ret}).
%
%\begin{equation}
%   \label{ret}
%\Phi(t, {\bf r})=\int{{\rho(\bar t, {\bf \bar r})\over{|{\bf r-\bar
%r|}}}}d\bar V\ ,
%\end{equation}
%
%with
%\begin{equation}
%   %\label{}
%\bar t=t-|{\bf r-\bar r|}\ ,
%\end{equation}
%
%
%\begin{equation}
%   %\label{}
%|{\bf r-\bar r|}=\sqrt {r^2-2{\bf r \bar r} + \bar r^2}\ ,
%\end{equation}
%
%$r=|{\bf r}|=(x^ix^i)^{1\over 2}$, similarly for ${\bf\bar r}$. 
Assume now general $\rho$ which is bounded in
space; then  at large distance from the matter, $r\gg \bar r$, we obtain
expansions for $\rho$ in (\ref{ret}).  

We  write
\begin{equation}
   \label{dif}
{|\bf r-\bar r|}=r+a(\bar r,\phi,\theta,\bar\phi,\bar\theta)+O\left({1\over r}\right)\ ,
\end{equation}
and
\begin{equation}
   \label{idif}
{1\over{|\bf r-\bar r|}}={1\over r}+O\left({1\over r^2}\right)\ .
\end{equation}
Because $\rho$ is helical we have
\begin{equation}
   %\label{}
\rho(\bar t,\bar\phi,\bar\theta,\bar r)=\hat\rho(\bar\phi-\Omega\bar
t,\bar\theta,\bar r)\ ,
\end{equation}
where $\hat\rho$ is periodic with period $2\pi$ in its first argument.
Hence we obtain
\begin{equation}
   %\label{}
\Phi(t,\phi,\theta,r)=\int{\hat\rho(\bar\phi-\Omega(t-{|\bf
r-\bar r|}),\bar\theta ,\bar r)\over{|\bf r-\bar r|}}d\bar V\ .
\end{equation}
Let us first investigate the field for $r\to\infty$, $t$ fixed. 
We can expand, using (\ref{dif}), 
%
%\begin{eqnarray}
%%\label{}
%&&\!\!\!\!\!\!\!\!\hat\rho\left(\bar\phi-\Omega(t-{|\bf r-\bar r|}),\bar\theta , \bar r\right) =
%\\\nonumber &&
%=
%\hat\rho\left(\bar\phi-\Omega(t-{r-a(\bar
%r,\phi,\theta,\bar\phi,\bar\theta)}),\bar\theta ,\bar r\right)
%\\\nonumber &&
%+{\partial\hat\rho\over\partial\bar\phi }\left(\bar\phi-\Omega(t-{r+a(\bar
%r,\phi,\theta,\bar\phi,\bar\theta)}),\bar\theta ,\bar r\right)
%\\\nonumber && \times O\left({1\over r}\right)+\dots  
%\end{eqnarray}
%
\begin{equation}
%\label{}
\hat\rho\left(\bar\phi-\Omega(t-{|\bf r-\bar r|}),\bar\theta , \bar r\right) = ~~~~~~~~~~~~~~~~~~~~~~~
\end{equation}
\vskip -8mm
$$
=
\hat\rho\left(\bar\phi-\Omega(t-{r-a(\bar
r,\phi,\theta,\bar\phi,\bar\theta)}),\bar\theta ,\bar r\right)~~~~~~
$$
\vskip -8mm
$$
+{\partial\hat\rho\over\partial\bar\phi }\left(\bar\phi-\Omega(t-{r+a(\bar
r,\phi,\theta,\bar\phi,\bar\theta)}),\bar\theta ,\bar r\right)
\times O\left({1\over r}\right)+\dots  
$$
(We can easily write a complete Taylor expansion if needed.)

 The leading
contribution to the field is therefore [see also (\ref{idif})] 
\begin{equation}
   %\label{}
\Phi(t,\phi,\theta,r)=~~~~~~~~~~~~~~~~~~~~~~~~~~~~~~~~~~~~~~~~~~~~~~~~~~~~\hfil
\end{equation}
\vskip -6mm
$$
=
{1\over
r}\int{\hat\rho\left(\bar\phi-\Omega(t-{r-a(\bar
r,\phi,\theta,\bar\phi,\bar\theta)}),\bar\theta ,\bar r\right)} d\bar V
+O\left({1\over r^2}\right)\ .
$$
The function
\begin{equation}
   %\label{}
P(t-r,\phi,\theta)=\int{\hat\rho\left(\bar\phi-\Omega(t-{r-a(\bar
r,\phi,\theta,\bar\phi,\bar\theta)}),\bar\theta ,\bar r\right)
d\bar V}
\end{equation}
is periodic in $t-r$ with period $2\pi/\Omega$ and the asymptotic behaviour is oscillatory in $r$ for
$t$ fixed, 
$r\to\infty$:
\begin{equation} 
   %\label{}
\Phi(t,\phi,\theta,r)\sim{P(t-r,\phi,\theta)\over r}\ .
\end{equation}
For a helical source with $Y_{lm}$ behaviour, 
\begin{equation}
   %\label{}
\hat\rho(\hat\phi,\theta,r)=e^{im\hat\phi}P_l(\cos\theta)f(r)\ ,
\end{equation}
we can see  the periodicity explicitly.

Next  we consider the asymptotic behaviour at future null infinity  $\mathcal{I^+}$.
With
$u=t-r$, we have  
\begin{eqnarray}
   %\label{}
&&\!\!\!\!\!\!\hat\rho\left(\bar\phi-\Omega(u+r-{|\bf r-\bar r|}),\bar\theta ,\bar r\right)=
\\\nonumber
&&=\hat\rho\left(\bar\phi-\Omega(u-a(\bar
r,\phi,\theta,\bar\phi,\bar\theta)),\bar\theta ,\bar r\right)+
\\\nonumber
&&
+{\partial\hat\rho\over\partial\bar\phi }\left(\bar\phi-\Omega(u-a(\bar
r,\phi,\theta,\bar\phi,\bar\theta)),\bar\theta ,\bar r\right)\times
O\left({1\over r}\right)+\dots \ . 
\end{eqnarray}
If we put this in the retarded integral (\ref{ret}) we find that for
$u=$const,
$r\to\infty$, we obtain  the usual Bondi--type  expansion 
in $r^{-1}$ (see
\cite{bo}, \cite{pen}) as we approach
$\mathcal{I^+}$.  

If we go to past null infinity 
$\mathcal{I^-}$, we have 
$v=t+r$ and obtain
%
%\begin{eqnarray}
%   %\label{}
%&&\!\!\!\!\!\!\hat\rho\left(\bar\phi-\Omega(v+2r-{|\bf r-\bar r|}),\bar\theta ,\bar
%r\right)=
%\\\nonumber&&
%\!\!\!=\hat\rho\left(\bar\phi-\Omega(v+2r-a(\bar
%r,\phi,\theta,\bar\phi,\bar\theta)),\bar\theta ,\bar r\right)+
%\\\nonumber&&
%+{\partial\hat\rho\over\partial\bar\phi }\left(\bar\phi-\Omega(v+2r-a(\bar
%r,\phi,\theta,\bar\phi,\bar\theta)),\bar\theta ,\bar r\right)
%\\\nonumber
%&&\times
%O\left({1\over r}\right )+\dots\ .  
%\end{eqnarray}
%
\begin{equation}
\hat\rho\left(\bar\phi-\Omega(v+2r-{|\bf r-\bar r|}),\bar\theta ,\bar
r\right)=~~~~~~~~~~~~~~~~~~~~~
\end{equation}
\vskip -8mm
$$
\!\!\!=\hat\rho\left(\bar\phi-\Omega(v+2r-a(\bar
r,\phi,\theta,\bar\phi,\bar\theta)),\bar\theta ,\bar r\right)+
$$
\vskip -8mm
$$
+{\partial\hat\rho\over\partial\bar\phi }\left(\bar\phi-\Omega(v+2r-a(\bar
r,\phi,\theta,\bar\phi,\bar\theta)),\bar\theta ,\bar r\right)
\times
O\left({1\over r}\right )+\dots\ .  
$$
This for $r\to\infty$, $v=$const implies   again   oscillations. This
behaviour is easy to understand in a spacetime picture: when going to 
$\mathcal{I^-}$ through the outgoing field of a source which has been periodically
moving at all times, we cross infinitely many oscillations.

\subsection{The Maxwell field}

We take a continuous compact distribution of charges   with
helical motion.  They define a conserved, helical 4--current. If
we use Lorentz gauge we have for each component of the
4--potential  to solve a scalar wave equation with a helical
source. Because the source is spatially compact the Lorentz
condition is satisfied for the solutions given by the retarded, or
advanced potentials. Hence, all the results for the scalar field
 apply.

\subsection{Linearized  gravity}
If we take the $T^{\mu\nu}$ of particles moving along the orbits
of a  helical Killing vector  we can consider  the linearized
equations in harmonic gauge. We then encounter again wave
equations for individual components and can write down retarded
and advanced solutions. The asymptotics is thus very similar to
that of the scalar fields. However, now  we cannot satisfy the
harmonicity condition because the energy momentum tensor in
linearized gravity for circular orbits   is not 
conserved.  A difference to the Maxwell case.

If we want a complete solution the matter must satisfy the equation of motion. For
example, those of linearized elasticity.

\section{The field of a point scalar charge in a circular orbit}
We adopt the standard procedure of finding retarded and advanced
Li\'{e}nard--Wiechert potentials in electrodynamics (see, e.g., an elegant,
covariant description by Rohrlich \cite{ro}) to the case of scalar
fields. The retarded ($\epsilon=+1)$ and advanced ($\epsilon=-1$)
solutions of the wave equation  with a
\  $\delta-$function type source with scalar charge $Q$ moving along the
worldline $\bar x^\mu(\tau)$ so that 
\begin{equation}{}
\rho(x^\mu)= Q\int\delta^{(4)}(x^\mu-\bar x^\mu(\tau))d\tau\ ,
\end{equation}
can be written in the form
\begin{equation}\label{sf}
\Phi_\epsilon=-{Q\over w_\epsilon}\ ,
\end{equation}
where 
\begin{equation}\label{37}
w_\epsilon= -\epsilon\ \eta_{\alpha\beta}{v^\alpha_\epsilon\over
c}R^\beta_\epsilon\ .
\end{equation}
(Henceforth, we do {\it not} set set $c=1$.)
Here the null vectors 
\begin{equation}\label{nv}
R^\beta_\epsilon=x^\beta-\bar x^\beta_\epsilon
\end{equation}
connect a given spacetime point $x^\beta$, in which
$\Phi_\epsilon$ is to be  calculated, with two points $\bar
x^\beta_\epsilon$ on the worldline $\bar
 x^\beta(\tau)$, in which the past and future light cones with vertex at
$x^\beta$ intersect the worldline. The 4--velocities $v^\alpha_\epsilon$
are evaluated at these points $\bar x^\beta_\epsilon$. The null vectors
 can be written as 
\begin{equation}\label{nv2}
R^\alpha_\epsilon=(\epsilon R=\epsilon|\bf R|,\bf R=\bf r- \bf{\bar
r}_\epsilon)\ ,
\end{equation}
the time components $R^0_\epsilon=\epsilon R$ are equal to the retarded and
minus advanced distance of the particle to the field point $x^\alpha$.

Now consider a point charge $Q$ moving in the plane $z=0$ along the circular
orbit of radius $a$ with center at $x=y=z=0$. So 
\begin{equation}\label{40}
{\bf{\bar r}}=a\left(\cos(\Omega t+\phi_0),\sin(\Omega t+\phi_0),0\right)\ ,
\end{equation}
where $\Omega$=constant is the particle's angular velocity, $\phi_0$ is the
azimuth at which the charge occurs at $t=0$. The 4--velocity is
\begin{equation}\label{41}
v^\mu=(c\gamma,\gamma\bf v)\ ,
\end{equation}
\begin{equation}{}
\gamma={1\over \sqrt{1-{a^2\Omega^2\over c^2}}}={\rm constant}\ ,
\end{equation}
the 3--velocity  is 
\begin{equation}\label{43}
{\bf v}=a\Omega \left(-\sin(\Omega t+\phi_0),\cos(\Omega t+\phi_0),0\right)\
.
\end{equation}
From these simple relations we find the distance $R$ entering the null
vectors (\ref{nv}) to be given by 
\begin{eqnarray}{}
R_\epsilon&=&\left(~(x-a\cos(\Omega t_\epsilon+\phi_0))^2\right.
\\\nonumber
&&\left.+(y-a\sin(\Omega
t_\epsilon+\phi_0))^2+z^2\right)^{1\over 2}\ ,
\end{eqnarray}
which, in spherical coordinates, implies
\begin{equation}{}
R_\epsilon=r \left(1-{2a\over r}\sin\theta\cos(\phi-\Omega
t_\epsilon -\phi_0)+{a^2\over r^2}\right)^{1\over 2}\ .
\end{equation}

Retarded and advanced times are implicitly given by the equation
\begin{eqnarray}\label{tra}
t_\epsilon&=&t-\epsilon{R_\epsilon\over c}
\\\nonumber
&=&t-\epsilon{r\over c}\left(1-{2a\over
r}\sin\theta\cos(\phi-\Omega t_\epsilon -\phi_0)+{a^2\over
r^2}\right)^{1\over 2}\ .
\end{eqnarray}
Note 
that on geometrical grounds we know that this equation has a unique solution for $t_\epsilon$
provided the worldline of the source is timelike, i.e., ${
a\Omega\over c}<1.$ In contrast to the case of linear uniform or
of uniformly accelerated motion, it is however impossible to
obtain an explicit expression for
$t_\epsilon$ as a function of $(t,r,\theta,\phi)$.

The expression in  (\ref{37}) for the scalar $w_\epsilon$ now reads
\begin{eqnarray}\label{w}
w_\epsilon=\gamma r&&\!\!\!\!\!\!\!\left(\left(1-{2a\over r}\sin\theta\cos(\phi-\Omega
t_\epsilon -\phi_0)+{a^2\over r^2}\right)^{1\over 2}-\right.
\nonumber\\
&&\left. -\epsilon
{a\Omega\over c}\sin\theta\sin(\phi-\Omega t_\epsilon-\phi_0)\right)\ ,
\end{eqnarray}
which then implies directly the form of the resulting scalar field
(\ref{sf}).

Let us finally rewrite the last relations in terms of the 'corotating '
angular coordinate $\hat\phi=\phi-\Omega t$. Introducing 
\begin{equation}{}
\hat\phi_\epsilon=\phi-\Omega t_\epsilon \ ,
\end{equation}
the equation for the retarded/advanced time becomes
\begin{equation}{}
\hat\phi_\epsilon=\phi+\epsilon {\Omega r\over c}\left(1-{2a\over
r}\sin\theta\cos(\hat\phi_\epsilon-\phi_0)+{a^2\over
r^2}\right)^{1\over 2}\ .
\end{equation}
Then the resulting scalar field,
\begin{eqnarray}{}
\Phi_\epsilon=-{Q\over \gamma r}&&\!\!\!\!\!\!\left\{\left(1-{2a\over
r}\sin\theta\cos(\hat\phi_\epsilon-\phi_0)+{a^2\over
r^2}\right)^{1\over 2}-\right.
\nonumber\\&&\left. -\epsilon{a\Omega \over
c}\sin\theta\sin(\hat\phi_\epsilon-\phi_0)\right\}^{-1}
\end{eqnarray}
is independent of the time $t$.

Let us finally investigate the asymptotic behaviour of
 this solution. Define the following dimensionless quantities
\begin{equation}\label{51}
\mathcal{C}_\epsilon=\sin\theta\cos(\phi-\Omega t_\epsilon -\phi_0)\ ,
\end{equation}
\begin{equation}\label{52}
\mathcal{S}_\epsilon=\sin\theta\sin(\phi-\Omega t_\epsilon -\phi_0)\ ,
\end{equation}
\begin{equation}{}
\mathcal{R}_\epsilon=\left( 1-2 \mathcal{C}_\epsilon {a\over r} +
{a^2\over r^2}   \right)^{1/2}\ ,
\end{equation}
\begin{equation}\label{54}
\alpha={a\Omega\over c}\ .
\end{equation}
For a subluminal source, $\alpha<1$; in the non--relativistic
limit,
$\alpha\ll 1$. At large $r\gg a$ we find 
\begin{equation}\label{asr}
\mathcal{R}_\epsilon=1- \mathcal{C}_\epsilon {a\over r} +
 {1\over 2}\mathcal{S}_\epsilon^2\ {a^2\over r^2} + O\left(\left({a\over
r}\right)^3 \right) \ ,
\end{equation}
irrespective of $t, t_\epsilon,\theta,\phi$.
This implies for the scalar field to be given at $r\gg a$ by the
expansion
\begin{equation}\label{sfa}
\Phi_\epsilon= - {Q\over \gamma r
(1-\epsilon\alpha\mathcal{S}_\epsilon)}\left(1+{\mathcal{C}_\epsilon\over
(1-\epsilon\alpha\mathcal{S}_\epsilon)}\ {a\over r}  
\right)+O\left(\left({a\over
r}\right)^3 \right)\ .
\end{equation}
For $\Omega=0$ we of course get the asymptotic static field of a source at
rest at $r=a,\phi=\phi_0, \theta =\pi/2$. 

For a moving source we have to take into account equation (\ref{tra}) for the
retarded/advanced time by employing the asymptotic form (\ref{asr}) for
$\mathcal{R}_\epsilon$:
\begin{equation}\label{aret}
t_\epsilon= t-\epsilon {r\over c} \mathcal{R}_\epsilon=t-\epsilon
{r\over c}+\epsilon{a\over c}\mathcal{C}_\epsilon
-\epsilon{1\over 2c}
\mathcal{S}^2_\epsilon \ {a^2\over r^2}+O\left({1\over r^2}\right)\ .
\end{equation}
(If we drop the $O$--term we obtain a new, approximate implicit
equation for
$t_\epsilon$.)
Fixing $t,\theta,\phi$, we get 

\begin{equation}{}
{\partial t_\epsilon\over\partial r}\bigg\vert_{t,\theta,\phi}=
{-\epsilon/c\over 1-\epsilon\alpha\mathcal{S}_\epsilon} +
O\left({1\over r}\right).
\end{equation}
Since $1-\epsilon\alpha\mathcal{S}_\epsilon>0$, $t_\epsilon$ is monotonically
decreasing ($\epsilon=1$) or increasing ($\epsilon=-1$) function of $r$.

Consider first the limit $t$ fixed, $r\to\infty$: The functions
$\mathcal{C}_\epsilon$, $\mathcal{S}_\epsilon$ are oscillatory as
$r\to\infty$ and, correspondingly, the scalar field decays, while oscillating,
in accordance with Eq. (\ref{sfa}). Retarded and advanced fields have similar
asymptotic behaviour at spatial infinity $i_0$.

To see the character of the oscillations in more detail, we solve
the equation for the dimensionless quantity $\Omega t_\epsilon$
iteratively (realizing that the term
$|\epsilon\alpha\mathcal{C}_\epsilon|<1$). So,
\begin{equation}\label{itr}
\Omega t_\epsilon\simeq \Omega t -\epsilon {\Omega r\over c}
+\epsilon\alpha\sin\theta\cos(\phi-\phi_0-\Omega(t-\epsilon{r\over c}))\ .
\end{equation}
The functions appearing in the fall--off of $\Phi_\epsilon$, with fixed
$t,\theta, \phi$ and $r\to\infty$, thus behave as 
\begin{equation}\label{60}
\mathcal{S}_\epsilon= p_1\sin\left(\epsilon{\Omega\over c} r+\epsilon
p_2\cos(\epsilon{\Omega\over c}r + p_3)+p_4\right)\ ,
\end{equation}
$p_1\dots p_4$ are constant parameters depending on fixed
$t,\theta, \phi$; similarly for $\mathcal{C}_\epsilon$. Notice
that $\mathcal{S}_\epsilon$ is again periodic in $r$ with the
period $2\pi c/\Omega$.

At null infinity, retarded and advanced fields behave very differently.
Turning to future null infinity $\mathcal{I}^+$, we put $u=t-{r\over c}$ and
consider the limit $r\to\infty$, with $u, \theta, \phi $ fixed.
From the asymptotic form of the equation (\ref{aret}) for
$t_\epsilon$ for $r\to\infty, u, \theta, \phi$ fixed we now find
that ${\partial t_{+}\over\partial r}=O(r^{-1})$,
whereas ${\partial t_{-}\over\partial r}={2\over
c}(1-\alpha\mathcal{S}_{-})^{-1}$. Here $\pm$ is the abbreviation for
$\epsilon=\pm$, i.e., retarded/advanced. Hence,
$t_{+}$ is approximately constant at large $r$, but $t_{-}$
is a monotonically increasing function of $r$. Indeed, using the
retarded time $t_{+}$ from (\ref{itr}) , we get
$\mathcal{C}_{+}, \mathcal{S}_{+}$ as functions
of $u,\theta, \phi$ , so that the retarded field given by
(\ref{sfa}) with $\epsilon=1$ has at $\mathcal{I}^+$ the standard
Bondi--type expansion \cite{bo}, \cite{pen}
\begin{equation}\label{61}
\Phi_{+}= -{Q\over\gamma}\left[{f_1(u,\theta,\phi) \over r} +
{f_2(u,\theta,\phi) \over r^2}  \right]+ O\left({1\over r^3}\right)\ .
\end{equation}
However, the advanced field---(\ref{sfa}) with $\epsilon=-1$---contains functions
like 
\begin{equation}\label{62}
\mathcal{S}_{-}=k_1\sin \left( -{2\over c}\Omega r + k_2\cos(-{2\over
c}\Omega r + k_3)+k_4
\right)\ ,
\end{equation}
which oscillate in $r$ with period ${\pi c/\Omega}$, so half of that at
spatial infinity.

At past null infinity $\mathcal{I}^-$, $r\to\infty$ with $v=t+{r\over c},
\theta,\phi$ fixed, the advanced fields exhibit the standard
Bondi--type expansion whereas retarded fields decay oscillating in
$r$. (All this becomes very clear in a spacetime picture.)

The oscillating factors at $r^{-k}$ in the asymptotic form (\ref{sfa}) of
$\Phi_\epsilon$ can become large for ultrarelativistic velocities when the
source moves near the light cylinder. 

For non--relativistic velocities, the
asymptotics of the fields simplifies considerably. Assuming $\alpha\ll 1$ and
neglecting terms which are $O(\alpha^2)$, we find the asymptotic expansion
(\ref{sfa}) to yield
\begin{eqnarray}\label{63}
\nonumber
\Phi_\epsilon&=& -{Q\over
r}\left\{1+\epsilon\alpha\sin\theta\sin\left(\phi-\phi_0-\Omega(t-\epsilon{r\over
c})\right)  
\right\}
\\\nonumber
&&-{Qa\over
r^2}\left\{1+\epsilon2\alpha\sin\theta\sin\left(\phi-\phi_0-\Omega(t-\epsilon{r\over
c})\right)  \right\}
\\
&&~~~~~\times\left\{\sin\theta\cos
\left(\phi-\phi_0-\Omega(t-\epsilon{r\over c})\right)\right\}\ .
\end{eqnarray}
From this simple  explicit expression the asymptotic behaviour of
$\Phi_\epsilon$ at  $i_0,\ \mathcal{I}^\pm$ becomes transparent. In
particular it is easy to see that at the leading order near $i_0$
both
$\Phi_\epsilon$ oscillate as $r^{-1}\sin(\epsilon\Omega r/c)$, whereas at
$\mathcal{I}^+$ the advanced field behaves like $r^{-1}\sin(2\Omega r/c)$ and,
in the same way, the retarded fields fall off at $\mathcal{I}^-$.

\section{The field of a electric point  charge in a circular
orbit}

As mentioned above, in the Lorentz gauge the results for the
individual components of the 4--potential will be essentially
the same as for the scalar field. It is, however, of interest to
see the behaviour of the electromagnetic field tensor because it
determines directly the force and, from the perspective of the
asymptotics, the peeling--off properties of its null tetrad
projections.  Since these may find applications also in other
context, we shall give the null--tetrad components in the whole
spacetime for a point charge moving along a circular orbit with,
in general, relativistic speed (and so emitting synchrotron
radiation).

We start from a general, covariant expression for the
retarded/advanced field tensor of an electric point charge  $e$
moving with a 4--velocity $v^\alpha$ and a 4--acceleration
$a^\alpha$ \cite{ro}. It is convenient to define a unit
spacelike vector $u_\alpha$ which is orthogonal to $v^\alpha$,
$u^\alpha u_\alpha=1$, $u^\alpha v_\alpha =0$, such that the
null vectors $R^\alpha_\epsilon=x^\alpha-\bar x^\alpha_\epsilon$
 (see Eqs. (\ref{nv}), (\ref{nv2})) can be written as 
\begin{equation}\label{64}
R^\alpha_\epsilon=w_\epsilon\left(u^\alpha_\epsilon+\epsilon
{v^\alpha_\epsilon\over c}\right)\ .
\end{equation}
Conversely,
\begin{equation}{}
w_\epsilon=\eta_{\alpha\beta} u^\alpha_\epsilon R^\beta_\epsilon
=-\epsilon \eta_{\alpha\beta}{v^\alpha_\epsilon\over
c}R^\beta_\epsilon \ , 
\end{equation}
where the last expression is identical with (\ref{37}). The
4--potential is simply
\begin{equation}\label{66}
A^\alpha_\epsilon={e\over c}{v^\alpha_\epsilon\over w_\epsilon}\
,
\end{equation}
which shows that it differs from the scalar field $\Phi_\epsilon$
given in (\ref{sf}) just by the replacement $Q\to -{e\over
c}v^\alpha$. The resulting expression for the Maxwell field tensor
reads:
\begin{eqnarray}
\label{max}
F^{\mu\nu}_\epsilon=&&{e\over cw^2_\epsilon}v^{[\mu}_\epsilon
u^{\nu]}_\epsilon
\\\nonumber
&+&
{e\over c^2 w_\epsilon}\left({1\over c} a^{[\mu}_\epsilon
v^{\nu]}_\epsilon- u^{[\mu}_\epsilon\left({1\over c}v_\epsilon^{\nu]}
a_\epsilon+\epsilon a^{\nu]}_\epsilon\right)
\right)\ ,
\end{eqnarray}
where the small square brackets denote antisymmetrization
without "$1\over 2$",
$a_\epsilon:=\eta_{\alpha\beta}a^\alpha_\epsilon
u^\beta_\epsilon$.

For the circular motion with the same parameters as with the
scalar charge all expressions for the velocity, retarded and
advanced times and the scalar $w_\epsilon$ are again given by
formulas (\ref{40})--(\ref{w}). In the expression (\ref{max}) we
also need the 4--acceleration  $a^\mu= dv^\mu/d\tau=\gamma
dv^\mu/dt$, where $v^\mu$ is given by  (\ref{41})--(\ref{43});
the result is 
\begin{equation}\label{68}
a^\mu_\epsilon=-{a\Omega^2\over {1-{a^2\Omega^2\over c^2}}}
\left(0,\cos(\Omega t_\epsilon+\phi_0),\sin(\Omega
t_\epsilon+\phi_0),0\right)\ .
\end{equation}
From (\ref{64}) the spacelike vector $u^\mu_\epsilon$ can be
expressed as  $u^\mu_\epsilon=R^\mu_\epsilon
{w_\epsilon}^{-1}-\epsilon v^\mu_\epsilon/c$, which in case
of the circular motion implies the components to be 
\begin{eqnarray}{}
u^0_\epsilon&=&\epsilon w^{-1}_\epsilon r
{R}_\epsilon-\epsilon \gamma\ , 
\\
u^1_\epsilon&=&w^{-1}_\epsilon\left(r\sin\theta\cos\phi-a\cos(\Omega
t_\epsilon+\phi_0)\right)+
\\\nonumber
&&+{\epsilon\over c}\gamma a \Omega \sin
(\Omega t_\epsilon+\phi_0))\ ,
\\
u^2_\epsilon&=&w^{-1}_\epsilon\left(r\sin\theta\sin\phi-a\sin(\Omega
t_\epsilon+\phi_0)\right)-
\\\nonumber
&&-{\epsilon\over c}\gamma a \Omega \cos
(\Omega t_\epsilon+\phi_0))\ ,
\\
u^3_\epsilon&=&w_\epsilon^{-1}r \cos \theta  \ ;
\end{eqnarray}
it is easy to check that $u^\mu_\epsilon$ is unit and
perpendicular to $v^\mu_\epsilon$. The last necessary ingredient
is the projection $a_\epsilon$ of $a^\mu_\epsilon$ on
$u^\mu_\epsilon$:
\begin{equation}
\label{pro}
a_\epsilon=-w^{-1}_\epsilon a^2\Omega^2 \gamma^2\left( {r\over
a}\sin\theta\cos(\phi-\Omega t_\epsilon-\phi_0)-1
\right)\ .
\end{equation}
The 4--potential is determined by (\ref{66}) in terms of the
quantities we know already from the scalar field case --- the
scalar $w_\epsilon$, see (\ref{w}), and by the 4--velocity
$v^\mu_\epsilon$ given by (\ref{41})--(\ref{43}) with $t\to
t_\epsilon$. Notice that the time--component $A^t_\epsilon$ is
equal to the scalar field $\Phi_\epsilon$ [see (\ref{sf})] up to
the sign and the multiplicative $\gamma$ factor. 

For highly relativistic motions, $A^t_\epsilon$ is thus much
bigger than the corresponding $\Phi_\epsilon$. The asymptotics
of $A^t_\epsilon$  is, however, the same as of $\Phi_\epsilon$
(see (\ref{sfa}), (\ref{61}) and (\ref{63})). Owing to the symmetry,
$A^z_\epsilon=0$. Because of the velocity entering the
componets $A^x_\epsilon$, $A^y_\epsilon$, their asymptotic
expansion at $i_0$ begins immediately with the oscillatory terms
at $r^{-1}$ even for non--relativistic motions $(\alpha\ll 1)$,
rather than with  constant$\times  r^{-1}$ as in (\ref{63}), and
the same with the advanced (retarded) fields at 
$\mathcal{I}^+$ ($\mathcal{I}^-$).

Substituting for $w_\epsilon,
v^\mu_\epsilon,a^\mu_\epsilon,u^\mu_\epsilon$ into (\ref{max})
we obtain all components of the Maxwell field tensor. These are
quite lengthy, here we give just the null tetrad components from
which all information can be retrieved. The standard null
(Newman--Penrose) tetrad (see, e.g., \cite{st}) in Minkowski spacetime
reads 
\footnote{This null tetrad is adapted to the spherical
coordinate lines rather than to Cartesian coordinates but its
components are Minkowskian expressed in terms of
$t,r,\theta,\phi$. The associated standard tetrad is given by
$e^\mu_{(0)}=(1/\sqrt2)(l^\mu+n^\mu),e^\mu_{(\hat
r)}=(1/\sqrt2)(l^\mu-n^\mu),e^\mu_{(\hat\theta)}=(1/\sqrt2)(m^\mu+\bar
m^\mu),e^\mu_{(\hat\phi)}=(i/\sqrt2)(m^\mu-\bar m^\mu)$. In spherical
coordinates,  $ e_{(\hat r)}=(0,1,0,0), e_{(\hat
\theta)}=(0,0,1/r,0),e_{(\hat\phi)}=(0,0,0,1/r\sin\theta)$.} 
\begin{eqnarray}
l^\mu &=& {1\over \sqrt 2}(1,
\cos\phi\sin\theta,\sin\phi\sin\theta, \cos\theta)\ ,
\nonumber\\\label{tetrad}
n^\mu &=& {1\over \sqrt 2}(1,
-\cos\phi\sin\theta,-\sin\phi\sin\theta, -\cos\theta)\ ,
\nonumber\\
m^\mu &=& {1\over \sqrt 2}(0,
\cos\phi\cos\theta -i\sin\phi,
\\\nonumber&&~~~~~~~~~~\sin\phi\cos\theta+i\cos\phi,
-\sin\theta)\ ;
\end{eqnarray}
in our signature it satisfies $l^\mu n_\mu=-1$, $m^\mu\bar
m_\mu=1$, all vectors are null.

Now the null tetrad electromagnetic {\it scalars}  are
given as the following projections of the Maxwell tensor:
\begin{eqnarray}\label{field}
\Phi_0&=&F_{\mu\nu}l^\mu m^\nu\ ,
\nonumber\\
\Phi_1&=&{1\over
2}F_{\mu\nu}(l^\mu n^\nu+\bar m^\mu m^\nu)\ ,
\\\nonumber
\Phi_2&=&F_{\mu\nu}\bar m ^\mu n^\nu\ \ .
\end{eqnarray}
The null tetrad naturally induces the orthonormal tetrad (see Ref. [18]) associated with the spherical coordinates with the standard
orthonormal triad $e_{\hat r}, e_{\hat\theta}, e_{\hat\phi}$
(supplemented by the timelike vector $(1,0,0,0)$). From the
scalars $\Phi_q, q=0,1,2$ the physical components of the
electric and magnetic field in the triad can be expressed by
means of the relation
\begin{eqnarray}{}
E_{\hat r}-iB_{\hat r}&=&2\phi_1\ ,
\nonumber\\
 E_{\hat\theta}-iB_{\hat
\theta}&=&-\Phi_0+\Phi_2\ ,
\\\nonumber
 E_{\hat\phi}-iB_{\hat
\phi}&=&-i(\Phi_0+\Phi_2)\ .
\end{eqnarray}
The Maxwell tensor field is given by (\ref{max}), with all the
ingredients expressed explicitly in (\ref{40})--(\ref{w}), 
(\ref{68})--(\ref{pro}). In addition to
$\mathcal C_\epsilon, \mathcal S_\epsilon, \mathcal R_\epsilon $ and
$\alpha$ determined by (\ref{51})--(\ref{54}), we introduce another
dimensionless quantity 
\begin{equation}{}
\mathcal Q_\epsilon=\mathcal R_\epsilon -\epsilon\alpha\mathcal
S_\epsilon=\left(1-2\mathcal C_\epsilon{a\over r}+{a^2\over
r^2}\right)^{1\over 2}-\epsilon\alpha\mathcal S_\epsilon \ .
\end{equation}
Finally, substituting the explicit expression for
$F^{\mu\nu}_\epsilon$ and the null tetrad (\ref{tetrad}) into
(\ref{field}) , we obtain --after straightforward though not very
short calculations and arrangements -- the following results for
the 3 complex scalars $\Phi_0, \Phi_1, \Phi_2$, in which again
$\hat\phi_\epsilon=\phi-\Omega t_\epsilon$:
\begin{widetext}

\begin{eqnarray}\label{phi0}
\Phi_0&=&{e\Omega\over{2cr\mathcal
Q^3_\epsilon}}\left\{-\alpha\left(1-\epsilon\mathcal
R_\epsilon -\mathcal C_\epsilon {a\over
r}+{a^2\over r^2}\right) \right.
\left[\cos\theta\cos(\hat\phi_\epsilon-\phi_0)-i\sin(\hat\phi_\epsilon-\phi_0)\right]
\\\nonumber
&&+(1-\epsilon\mathcal R){a\over
r}\left[-\cos\theta\sin(\hat\phi_\epsilon-\phi_0)-i\cos(\hat\phi_\epsilon-\phi_0)\right]
\left.
+i\left[\alpha^2(1-\epsilon\mathcal
R_\epsilon -\mathcal C_\epsilon {a\over
r})+{a^2\over r^2}\right]\sin\theta\right\}+
\\\nonumber
&&+{ea\over 2(r\mathcal
Q_\epsilon)^3}\left[\cos\theta\cos(\hat\phi_\epsilon-\phi_0)-i\sin(\hat\phi_\epsilon-\phi_0)\right]\
,
\\
\label{phi1}
\Phi_1&=&{e\over{2r^2\mathcal
Q^3_\epsilon}}\left\{1-\epsilon\alpha \mathcal
R_\epsilon\sin\theta\sin(\hat\phi_\epsilon-\phi_0)\right.
%\\\nonumber&&
-\left (\alpha^2\mathcal C_\epsilon+(1-\alpha^2){a\over
r}+i\alpha^3\cos\theta\right)\sin\theta\cos(\hat\phi_\epsilon-\phi_0)
%\\\nonumber&&
\left.
+i\alpha{a\over r}\cos\theta\right\}\ ,
\\
\label{phi2}
\Phi_2&=&-{e\Omega\over{2cr\mathcal
Q^3_\epsilon}}\left\{-\alpha(1+\epsilon\mathcal
R_\epsilon -\mathcal C_\epsilon {a\over
r}+{a^2\over r^2})
%\times
\right.
%\\\nonumber&&
%\times
\left[\cos\theta\cos(\hat\phi_\epsilon-\phi_0)+i\sin(\hat\phi_\epsilon-\phi_0)\right]
\\\nonumber&&
+(1+\epsilon\mathcal R_\epsilon){a\over
r}\left[\cos\theta\sin(\hat\phi_\epsilon-\phi_0)-\right.
%\nonumber\\&&
\left.i\cos(\hat\phi_\epsilon-\phi_0)\right]
\left.
+i\sin\theta\left[\alpha^2(1+\epsilon\mathcal
R_\epsilon -\mathcal C_\epsilon {a\over
r})+{a^2\over r^2}\right]\right\}
\nonumber\\&&
-{ea\over 2(r\mathcal
Q_\epsilon)^3}\left[\cos\theta\cos(\hat\phi_\epsilon-\phi_0)-i\sin(\hat\phi_\epsilon-\phi_0)\right]\ .
\end{eqnarray}
\vskip 2cm
\end{widetext}
There are well--known properties of the null--tetrad
components of the zero--rest mass fields (see, e.g.,
\cite{pen}). In the electromagnetic case $\Phi_2$
characterizes the outgoing radiation field at
$\mathcal{I}^+$: it is given as
$\lim_{r\to\infty,u=const}(r\Phi_2)$. At 
$\mathcal{I}^-$\ \  
$\lim_{r\to\infty,v=const}(r\Phi_0)$ determines the
incoming radiation field. We shall see that, indeed,
this is the case for $\Phi_{2+}$ at  $\mathcal{I}^+$ 
and for $\Phi_{0-}$ at  $\mathcal{I}^-$.  (Here
$\ \pm$ is again the abbreviation for $\epsilon=\pm 1$, i.e.,
retarded/advanced fields.)

%\bigskip

Before turning to the asymptotic properties of the
$\Phi$s  let us observe some intriguing "symmetry
relations" of these quantities. They arise as a
consequence of the helical symmetry of the source and
the fields. In order to understand  them we have to
turn back to the equation (\ref{tra}) for retarded and
advanced times. Notice first that putting the 'field'
time $t=0$, Eq.(\ref{tra}) still cannot be solved
explicitly for $t_\epsilon$, however it is easy to see
the following relations between $t_+$ and $t_-$:
\begin{equation}\label{ts}
t_+(t=0,r,\theta,\phi)=-t_-(t=0,r,\theta,-\phi +2\phi_0)\ .
\end{equation}
(It is instructive to draw the circular orbit of the
source and convince oneself that this relation can
also be understood on "geometrical grounds".) Owing to
the helical symmetry the relation (\ref{ts}) can
be generalized for any $t$ to read
\begin{equation}\label{tsg}
t_+(t,r,\theta,\phi+\Omega t)-t=-t_-(t,r,\theta, -\phi+\Omega t
+2\phi_0)+t \ ;
\end{equation}
this can be checked directly using (\ref{tra}) again.
 
%\bigskip
Now as a consequence of helical symmetry and, in particular, of (\ref{tsg}), 
there arise the following relations---valid at any spacetime 
point---between the retarded  and advanced fields as given
by the null--tetrad components of the $\Phi$'s:
\begin{equation}\label{tes}
\Phi_{2+}(t,r,\theta,\phi+\Omega
t)=-\Phi_{0-}(t,r,\theta,-\phi+\Omega t+2\phi_0)\ ,
\end{equation}
$$
\Phi_{0+}(t,r,\theta,\phi+\Omega
t)=-\Phi_{2-}(t,r,\theta,-\phi+\Omega t+2\phi_0)\ ,
$$
$$
\Phi_{1+}(t,r,\theta,\phi+\Omega
t)=+\Phi_{1-}(t,r,\theta,-\phi+\Omega t+2\phi_0)\ .
$$
It is straightforward to convince oneself that our
resulting expressions (\ref{phi0})--(\ref{phi2})
satisfy these relations. They are the keystone in
Schild's 2--body problem, in fact the $N$--body as well,
implying that tangential forces on particles vanish
when both retarded and advanced effects are taken into
account. But this problem will be considered elsewhere
(for the $N$--particle problem in the case of the scalar
field, see \cite{bhs}).

%\bigskip
We now turn to the asymptotic properties of the
fields. As with the scalar field we first make the
expansions of the retarded and advanced fields at
large $r\gg a$ using the expansion (\ref{asr}) of
$\mathcal R_\epsilon$.  In addition to the the abbreviations $\mathcal
C_\epsilon,\mathcal S_\epsilon$ defined in (\ref{51}), (\ref{52}), 
i.e.,
\begin{eqnarray}{}
\mathcal C_\epsilon&=&\sin\theta\cos(\hat\phi_\epsilon-\phi_0)\ ,
\\
\mathcal S_\epsilon&=&\sin\theta\sin(\hat\phi_\epsilon-\phi_0)\ ,
\end{eqnarray}{}
we introduce

\begin{eqnarray}
\tilde{\mathcal C}_\epsilon&=&\cos\theta\cos(\hat\phi_\epsilon-\phi_0)\ ,
\\
\tilde{\mathcal S}_\epsilon&=&\cos\theta\sin(\hat\phi_\epsilon-\phi_0)\ ,
\end{eqnarray}
and
\begin{eqnarray}
c_\epsilon&=&\cos(\hat\phi_\epsilon-\phi_0)\ ,
\\
s_\epsilon&=&\sin(\hat\phi_\epsilon-\phi_0)\ .
\end{eqnarray}

With these notations we  obtain
the following: 
\begin{widetext}
\begin{eqnarray}%\label{}
\Phi_{2+}&=&{e\Omega\alpha\over c(1-\alpha {\mathcal S}_+)^3}\left(
\tilde{\mathcal C}_++is_+
-i\alpha\sin\theta
\right) 
{1\over r}+O\left({1\over r^2}\right)\ ,
\\\nonumber
\Phi_{2-}&=&-{ea\over 2(1+\alpha\mathcal
S_-)^3}
\{\tilde{\mathcal C}_-+is_-
+\alpha[C_-(\tilde{\mathcal S}_--ic_-)+
\\
&&+i\sin\theta-{1\over2}\alpha(1+{\mathcal C}^2_-)(\tilde{\mathcal C}_-+is_-)-i{\alpha^2\over 2}
(1-{\mathcal C}^2_-)\sin\theta]\}
{1\over r^3} + O\left({1\over r^4}\right)\ ,
\\
\Phi_{1\epsilon}&=&{e\over 2(1-\epsilon\alpha
{\mathcal S}_\epsilon)^3}\left[ 1-\epsilon
\alpha{\mathcal S}_\epsilon-(\alpha^2{\mathcal C}_\epsilon
+i\alpha^2\cos\theta){\mathcal C}_\epsilon\right]
{1\over r^2}+ O\left({1\over r^3}\right)\ ,
\\\nonumber
\Phi_{0+}&=&-{ea\over 2(1-\alpha\mathcal
S_+)^3}
\{\tilde{\mathcal C}_+-is_+ +\alpha[-{\mathcal C}_+(\tilde{\mathcal S}_+-ic_+)+
\\
&&+i\sin\theta-{1\over2}\alpha(1+{\mathcal C}^2_+)(\tilde{\mathcal C}_+-is_+) 
- i{\alpha^2\over 2}(1-{\mathcal C}^2_+)\sin\theta]\}
{1\over r^3}+O\left({1\over r^4}\right)\ ,
\\
\Phi_{0-}&=&-{e\Omega\alpha\over c(1+\alpha{\mathcal S}_-)^3}
\left( \tilde{\mathcal C}_--is_-
-i\alpha\sin\theta \right) 
{1\over r}+O\left({1\over r^2}\right)\ .
\end{eqnarray}
\vskip1cm
\end{widetext}

\bigskip
These formulas get still simplified for
nonrelativistic motion of the source, $\alpha\ll1$,
when only terms linear in $\alpha$ are kept (using
$(1-\epsilon\alpha\mathcal S_\epsilon)^{-3}\simeq
1+3\alpha\mathcal S_\epsilon+O(\alpha^2))$ and
rearranging the terms:
\begin{equation}\label{90}
\Phi_{2+}={e\Omega\alpha\over
c}\left(\tilde{\mathcal C}_++is_+\right){1\over r}+O\left({1\over
r^2}\right)\ ,
\end{equation}
\begin{equation}{}
\Phi_{2-}=-{ea\over
2}\left[(1-2\alpha {\mathcal S}_-)(\tilde {\mathcal C}_-+is_-)\right]{1\over
r^3}+O\left({1\over r^4}\right)\ ,
\end{equation}
\begin{equation}{}
\Phi_{1\epsilon}={e\over 2}(1+2\epsilon\alpha
{\mathcal S}_\epsilon){1\over r^2}+O\left({1\over r^3}\right)\ ,
\end{equation}
\begin{equation}{}
\Phi_{0+}={ea\over
2}\left[(1+2\alpha \mathcal
S_+)(\tilde{\mathcal C}_+-is_+)\right]{1\over r^3}+O\left({1\over
r^4}\right)\ ,
\end{equation}
\begin{equation}\label{94}
\Phi_{0-}=-{e\Omega\alpha\over
c}\left(\tilde{\mathcal C}_--is_-\right){1\over r}+O\left({1\over
r^2}\right)\ .
\end{equation}

Clearly, the fields have again the oscillatory
character. The retarded and advanced times
$t_\epsilon$ can be expressed in terms of the field
time $t$ and functions $\mathcal C_\epsilon$ and
$\mathcal S_\epsilon$ according to Eq. (\ref{aret}).
At fixed $t,\theta,\phi$ and $r\to\infty$, i.e., at
$i_0$, the oscillations are given by $\mathcal
C_\epsilon$, $\mathcal
S_\epsilon$ which have the form (\ref{60})
  as in the scalar--field case.
The retarded and advanced fields behave similarly.

\bigskip
At future null infinity $\mathcal I^+$ the retarded
field has the standard Bondi--type expansion since $\mathcal
C_+$, $\mathcal
S_+$ become functions of $u,\theta,\phi$ [see
Eq. (\ref{61}) and the text therein]. In addition, in the
electromagnetic case we observe a new feature in the
asymptotics -- the well--known "peeling--off property"
\cite{pen}, \cite{st}: $\Phi_{q+}\sim
f_q(u,\theta,\phi)r^{q-3}$. The limits
$r^{3-q}\Phi_{q+}$ are well defined at $r\to\infty$ with
$u=t-r/c, \theta,\phi$ fixed, i.e., at $\mathcal I^+$.
With the advanced field at $\mathcal I^+$ the situation
is different. We still find a "generalized peeling" in
the sense that $\Phi_{q-}\sim r^{-1-q}$; so $\Phi_{q-}$
corresponds to $\Phi_{(2-q)+}$. However, the limits
$r^{1+q}\Phi_{q-}$ at $r\to\infty,u,\theta,\phi$ fixed
do not exist since these are oscillatory functions as
$\mathcal S_-$ in (\ref{62}), i.e.,  with period
$\pi c/\Omega$ which is half of that at spatial
infinity. Nevertheless, notice that $r^{1+q}\Phi_{q-}$
remains bounded as $r\to\infty$.

\bigskip
As expected, at past null infinity $\mathcal I^-,
r\to\infty, v=t+r/c,\theta,\phi$ fixed, the advanced
fields exhibit the standard Bondi-type expansion and
peeling, whereas the retarded fields do decay with
$r\to\infty$ but in an oscillatory manner.

All these features become evident especially when
regarding the simple fields (\ref{90})--(\ref{94}) in the
non--relativistic limit $\alpha\ll 1$.

\section{Acknowledgments}

We thank J. Winicour for many useful discussions. 
J.B. gratefully acknowledges the support of the Alexander von
Humboldt Foundation and also  of the  grants No. GACR 202/06/0041,
No. LC06014 and MSM0021620860 of  the Czech Rebublic.


\begin{thebibliography}{99}


\bibitem{li}{L. Lichtenstein}, {\it Mathematische
Zeitschrift} {\bf 12}, {201} (1922).



\bibitem{de}{ J. K. Blackburn and S. Detweiler},  {\it Phys. Rev. D} {\bf 46},  2318 (1992).
{S. Detweiler}, {\it Phys. Rev. D} {\bf 50},  4929 (1994).


\bibitem{sch}{A. Schild},  {\it Phys.Rev.} {\bf 131(6)} ,
{2762} (1963).

\bibitem  {bhs} {R. Beig, J. M. Heinzle,  and B. G. Schmidt},
{\it Phys. Rev. Lett.}, to be published (2007).



\bibitem{fr}{J. L. Friedman, K. Ury\={u}, and M. Shibata},  {\it
Phys. Rev. D} {\bf 65}, 064035 (2002); {\bf 70}, 129904(E) (2004); 

{J. L. Friedman and K. Ury\={u}}, {\it Phys. Rev. D} {\bf 73},
104039 (2006);

{S. Yoshida, B. C. Bromley, J. S. Read, K. Uryu, and J. L.
Friedman},  {\it Class. Quantum Grav. }{\bf 23},
S599 (2006).


\bibitem{wh}{J. T. Whelan, W. Krivan, and R. H. Price},  {\it
Class. Quantum Grav.} {\bf 17}, 4895 (2000);

{B. Bromley, R. Owen, and R. H. Price}, {\it Phys. Rev. D} {\bf 71},
104017 (2005);

{C. Beetle, B. Bromley, and R. H. Price}, {\it Phys. Rev. D} {\bf
74},  024013 (2006).  

\bibitem{gou}{E. Gourgoulhon, P. Grandclement, and S. Bonazzola},
{\it  Phys. Rev. D} {\bf 65}, 044020 (2002);

{T. Damour, E. Gourgoulhon, and P. Grandclement}, {\it Phys. Rev.
D} {\bf 66}, 024007 (2002);

{E. Gourgoulhon, P. Grandclement, and S. Bonazzola},
{\it  Int. J. Mod. Phys.} A{\bf 17}, 2689 (2002).

\bibitem{kl}{C. Klein}, {\it Phys. Rev. D} {\bf 70}, 124026 (2004).

\bibitem{br}{J. P. Bruneton}, gr--qc/0611021, to appear in the
proceedings of Albert Einstein's Centunary International
Conference, Paris, France, 18--22 July;

{J. P. Bruneton and G. Esposito--Far\`{e}se (unpublished)}.

\bibitem{tor}{C. G. Torre}, {\it J. Math. Phys.} {\bf 44}, 6223
(2003); {\it J. Math. Phys.} {\bf 47}, 073501 (2006).



\bibitem{bo}{H. Bondi, M. G. J. van der Burg, and A. W. K.
Metzner},  {\it  Proc. Roy. Soc. Lond.} {\bf A269}, 21 (1962).

\bibitem{pen}{R. Penrose},  {\it Proc. Roy. Soc. Lond.} {\bf
A284}, 159 (1965).

\bibitem{wa}{R. M. Wald}, {\it General Relativity} (Univ. Chicago
Press, Chicago, 1984).

\bibitem{jack}{J. D. Jackson}  {\it Classical
electrodymics, 2nd Edition}  (John Wiley, New York 1975). 

 \bibitem{ab}{M. Abramowitz and I. A. Stegun}, {\it Handbook of
Mathematical Functions} (Dover Publ., New York 1972).

\bibitem{ro}{F. Rohrlich}, {\it Classical Charged Particles}
(Addison--Wesley, Reading, 1996).

\bibitem{st}{J. Stewart}, {\it Advanced general relativity}
(Cambridge University Press, Cambridge, U.K., 1991).

\end{thebibliography}
\end{document}